\newcommand{\Fig}[1]{Fig.~\ref{Fig:#1}}
\newcommand{\Eq}[1]{Eq.~(\ref{#1})}  
\newcommand{\ket}[1]{|#1\rangle}
\newcommand{\bra}[1]{\langle #1|}
\newcommand{\brkt}[2]{\langle #1 | #2 \rangle}

\newcommand{\A}{{\sf A}}
\newcommand{\B}{{\sf B}}
\newcommand{\op}[1]{\hat{#1}}
\newcommand{\Proj}{\op{\mathtt P}}
\newcommand{\Id}{\mathtt{1\!\!I}}
\newcommand{\Meas}{\op{\mathtt M}}
\newcommand{\Swap}{\op{\mathtt X}}
\newcommand{\Ket}[1]{\mathop{\left|#1\right\rangle}}
\newcommand{\mbs}[1]{\boldsymbol{#1}} 
\newcommand{\comp}{\ominus}
\newcommand{\JJ}{\pmb{\mathbb{J}}}
\newcommand{\pix}[3]{\raisebox{#1\height}{\includegraphics[scale=#2]{#3}}}
\newcommand{\sop}[1]{\breve{#1}}
\newcommand{\SPr}{\sop{\mathtt{\Pi}}}
\newcommand{\MesPar}[2]{(\!(#1,#2)\!)}
\newcommand{\Av}[1]{\langle #1\rangle}
\newcommand{\Abs}[1]{\bigl|#1\bigr|}
\newcommand{\newpart}{\bigskip}

\documentclass[12pt]{article}
\usepackage{amsmath,amssymb,exscale,graphicx}
\DeclareMathOperator{\Tr}{Tr} 

\numberwithin{equation}{section}
\title{\bfseries Quantum Mechanics and Nonlocality:\\ 
In Search of Instructive Description}
\date{\today}
\author{\em\Large Alexander \t{Y}u. Vlasov\thanks{%
Electronic mail: \protect\raisebox{-3pt}{\protect\includegraphics{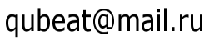}}}}
\begin{document}
\maketitle

\begin{abstract}
A problem with an instructive description of measurement process for 
sufficiently separated entangled quantum systems is well known.
More precise and crafty experiments together with new technological 
challenges raise questions about sufficiency of formal 
use of ``black-box'' Copenhagen paradigm without subtleties of 
transition between quantum and classical worlds.
In this work are discussed applications both standard interpretation of
quantum mechanics and ``unconventional'' models, like relative 
state formulation, multiple clocks formalism, and extended probabilities.
\end{abstract}

\section{Introduction}
\label{Sec:intro}



\begin{figure}[htb]
\begin{center}
\includegraphics[scale=0.75]{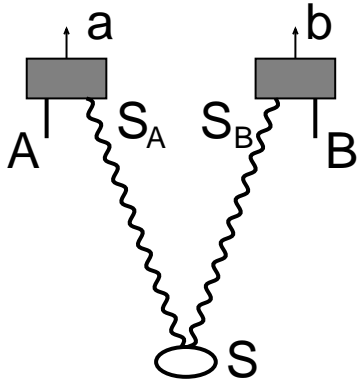} 
\end{center}
\caption{Scheme of measurement}\label{Fig:nonloc}
\end{figure}

There are certain problems with description of nonlocality in quantum mechanics.
A ``black-box'' scheme of a standard experiment is represented on \Fig{nonloc}: 
initially interacting quantum systems are separated on some distance and 
two independent measurements are performed after that.  

In a conclusion to a description of an experimental test of ``realism'' conception
in quantum mechanics, performed not so long time ago \cite{nonl} was mentioned, 
that these ``results lend strong support to the view that any future extension 
of quantum theory that is in agreement with experiments must abandon certain 
features of realistic descriptions.'' Rather broad notion of the realistic 
description may be found in very beginning of the same paper \cite{nonl}:
``Physical realism suggests that the results of observations are a consequence 
of properties carried by physical systems.'' 

It looks like suggestion for any future extension of quantum theory to evade
ideas about results of observation as a consequence of properties carried
by physical systems. 
The claim about limitations on ``realism'' considered in \cite{nonl}
is more strong, than in most other works, because it is ``experimentally excluded''
both local models and ``a class of important non-local hidden-variable theories.'' 
It is also mentioned, that non-local models outside of this class are 
``highly counterintuitive.''

On the other hand, yet another experiment suggested in \cite{CK,CK2v2,CK3} 
may be considered as an illustration, that paradox of quantum
nonlocality may be irrelevant with discussions about reality
of wave functions or density matrixes.
The Conway-Kochen Free State (Free Will) theorem \cite{CK} 
is formulated about results of observations and so illustration
of a problem is transferred to classical level.

Formally, it even does not matter for SPIN and TWIN axioms \cite{CK}, 
if the measurements are a consequence of quantum or classical properties 
carried by physical systems or not.
Only FIN axiom about finite speed of information transfer \cite{CK} or
MIN axiom about locality \cite{CK3} include some suggestions about 
properties of physical processes.

Problem with locality demonstrated by Conway-Kochen theorem is not 
produced by realism or any other property of physical model. An advantage 
of such result declared in \cite{CK} ``is that it applies directly to the 
real world rather than just to theories. It is this that prevents the 
existence of local mechanisms for reduction.'' 

In order to avoid discussions about free will \cite{CK,CK2v2,CK3,GtH,PerW} 
it could be suggested few explanations:
\begin{enumerate}
\setlength{\itemsep}{0pt}
\item Nonlocality.
\item Some effects ({\em e.g.}, relativistic, stochastic, {\em etc.}) 
obstructing possibility to use abridged version of quantum theory
expressed by SPIN, TWIN postulates \cite{CK}.  
\item Inapplicability of suggestion about definite outcomes ({\em e.g.}, theories
without reduction like Everett formulation).
\end{enumerate}

So, from the one hand, there are ideas about necessity to abandon realism to
have an agreement with experiment \cite{nonl}, but from the other hand
there is no any guarantee about resolution of problems with consistent
description of more complicated tests \cite{CK}.
It could be even said informally, that if classical notion of realism ``is  
not enough'' to establish interdependences displayed by quantum systems, 
then the idea to ``abandon certain features of realistic descriptions'' may be 
a step in a wrong direction.

The purpose of presented excursus was not a criticism of some particular ideas, 
but demonstration of rather objective difficulties in modern interpretations of
experiments with a {\em few} quantum systems. Nowadays quantum physics
displays enormous progress in predicting and explaining of properties of
elementary quantum particles, but already description of two simplest 
quantum systems may raise conceptual problems illustrated above.

\paragraph*{Contents:}

In Sec.~\ref{Sec:Nonloc} is reminded simple nonlocal model 
of measurement for arbitrary quantum system with finite-dimensional 
space of states. Nowadays such kind of models ensures possibility 
to use classical computer for modeling of not very large quantum 
systems in quantum information science. Problem with nonlocality
may be formally avoided in formulations without collapse and
it is discussed in Sec.~\ref{Sec:Ev}. Particular example of 
entangled spin-1 particles used in Conway-Kochen theorem is 
developed for the ``no-collapse'' formulation in App.~\ref{App:CKwor}.

Sec.~\ref{Sec:twot} is devoted to introduction of two-time 
(or, rather, two-clocks) formalism earlier suggested by Bell 
for description of EPR paradox and GRW models. 
Von Neumann measurement scheme is applied further in 
Sec.~\ref{Sec:vnNm} and problem of definite outcomes
is revisited in Sec.~\ref{Sec:defout}. This problem is
especially illustrative for Conway-Kochen scheme 
with two entangled spin-1 particles discussed
in Sec.~\ref{Sec:CKT} and App.~\ref{App:CKEx}.

Feynman ideas about negative probabilities are affected in
Sec.~\ref{Sec:negprob} and App.~\ref{App:Decomp} with
concrete example of decomposition with positive and
negative terms for entangled state of two spin-1 particles. 
Finally, in Sec.~\ref{Sec:insep} is touched upon a relation 
of nonlocality and inseparability in quantum mechanics.

\section{Simple Nonlocal Model}
\label{Sec:Nonloc}

A simple nonlocal hidden variable model for discussed experiments is known and
may be even considered as direct consequence of quantum mechanical definitions.

\begin{figure}[htb]
\begin{center}
\includegraphics[scale=0.75]{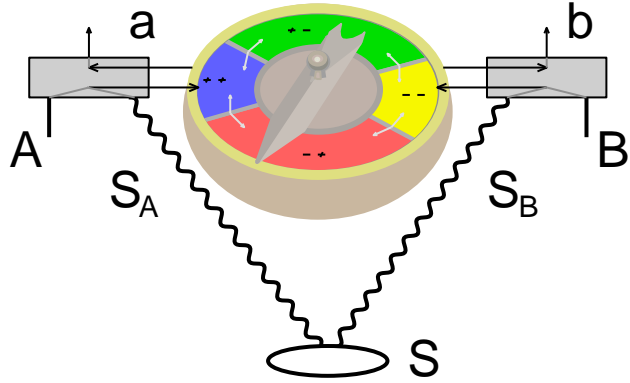} 
\end{center}
\caption{Nonlocal hidden variables model}\label{Fig:nonlochv}
\end{figure}

Let us discuss briefly such a model for experiments with two possible outcomes
of each measurement like \cite{nonl}. If there are known equations for probabilities
($P_1 \equiv P_{++}$, $P_2 \equiv P_{+-}$, $P_3 \equiv P_{-+}$, $P_4 \equiv P_{--}$)
for four possible combinations of events like $\sf a=\pm 1$ and $\sf b = \pm 1$ we
may use ``roulette'' \Fig{nonlochv} with four sectors proportional to given 
probabilities and angle of rotation of pointer is corresponding to a hidden 
variable $0 \le \phi < 2 \pi$. Pair of outcomes is defined by pairs
of indexes $(+1,+1)$, $(+1,-1)$, $(-1,+1)$, $(-1,-1)$
assigned with each sector of the roulette. 

It is usual example of a model for generation of given probability distribution
for known probabilities and so it may be considered as rather ``mechanical''
consequence of Born rules with squares of modules of quantum amplitudes.
But this ``minimal'' model is really counterintuitive for separated systems
because for application of such ``nonlocal roulette'' it is necessary to know 
parameters of measurements for {\em both} parties \A\ and \B.

In general case there are set of probabilities $P_{\sf ab}$ defined for
entangled state $\Psi$ and projectors $\Proj_{\sf a}$ and $\Proj_{\sf b}$
with equation like
\begin{equation}
 P_{\sf ab} = \bra{\Psi} \Proj_{\sf a} \otimes \Proj_{\sf b} \ket{\Psi}.
\label{Pab}
\end{equation}

For a product state $\ket{\Phi} = \ket{\psi}\ket{\phi}$ it is possible
to write
\begin{equation}
 P_{\sf ab} = \bra{\Phi} \Proj_{\sf a} \otimes \Proj_{\sf b} \ket{\Phi} =
 \bra{\psi}\Proj_{\sf a}\ket{\psi} \bra{\phi}\Proj_{\sf b}\ket{\phi} = P_{\sf a}P_{\sf b},
\quad P_{\sf a} = \bra{\psi}\Proj_{\sf a}\ket{\psi},
~P_{\sf b} = \bra{\phi}\Proj_{\sf b}\ket{\phi}.
\label{dePab}
\end{equation}
In classical probability theory \Eq{dePab} $P_{\sf ab} = P_{\sf a}P_{\sf b}$ corresponds 
to independent events and may be modeled with two local roulettes.

\begin{figure}[htb]
\begin{center}
a)\includegraphics[scale=0.5]{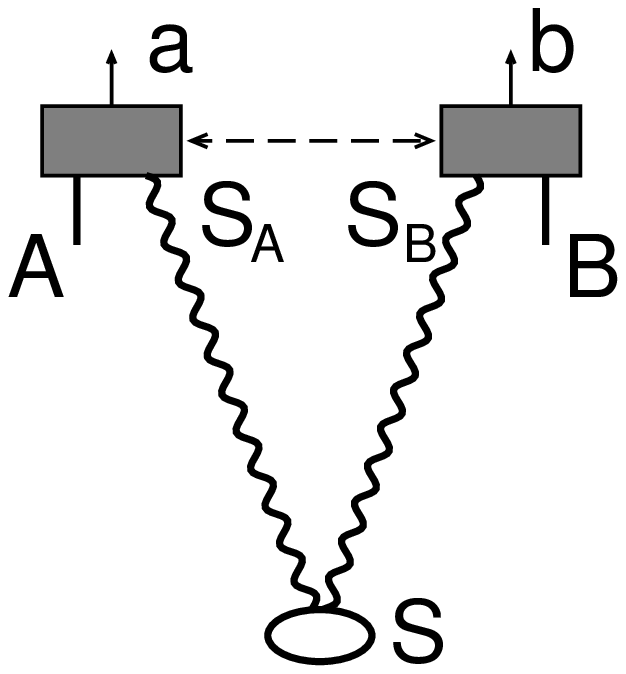} 
b)\includegraphics[scale=0.5]{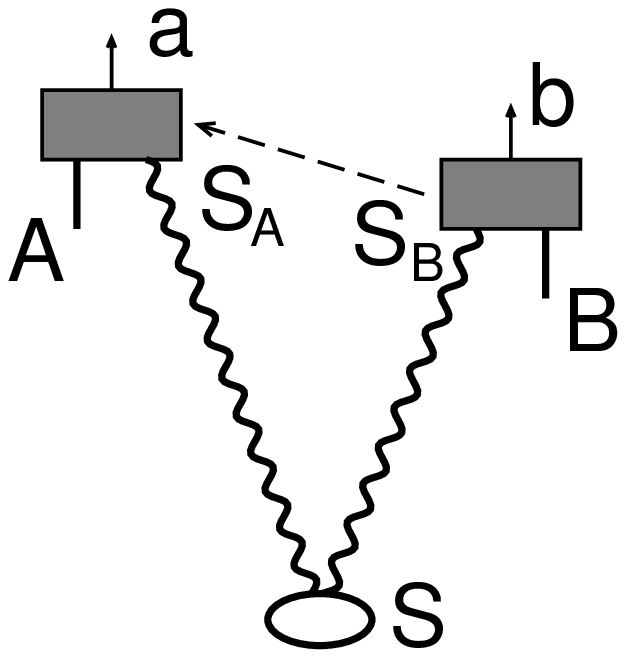}
c)\includegraphics[scale=0.5]{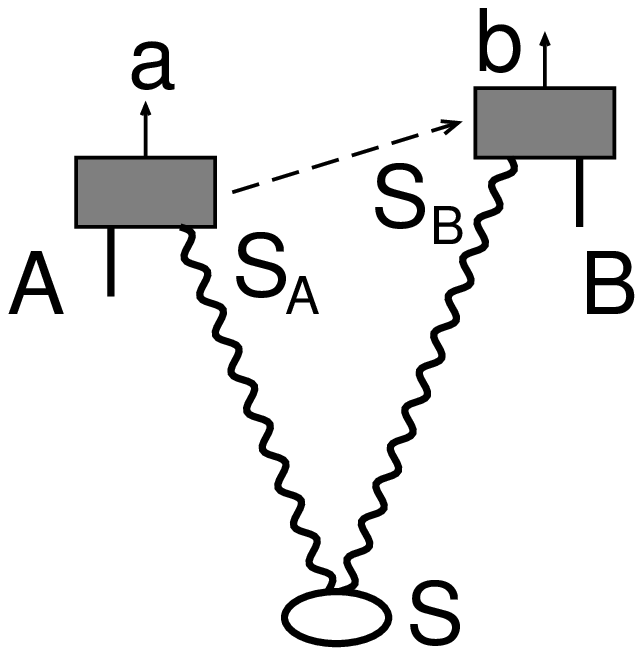}
\end{center}
\caption{Schemes of possible dependencies}\label{Fig:depend}
\end{figure}

It is also possible to reproduce correlated events, if to use analogue of classical
equation for conditional probabilities \cite{Fell}
\begin{equation}
\textstyle
 P(a,b) = P(a)P(b|a) = P(b)P(a|b), 
\label{condprob}
\end{equation}
and instead of model with two correlated events $P(a,b) \equiv P_{\sf ab}$ \Eq{Pab}  
(\Fig{depend}a), one event is considered as local one with probability $P_{\sf a}$ 
and the second one is dependent event with conditional probability
$P_{\sf b|a} = P_{\sf ab}/P_{\sf a}$ (\Fig{depend}c). It is also possible to use 
an opposite order of events with probabilities $P_{\sf b}$ and 
$P_{\sf a|b} = P_{\sf ab}/P_{\sf b}$  (\Fig{depend}b). Equations for  
$P_{\sf a}$ and $P_{\sf b}$ for given entangled state $\ket{\Psi}$ may be 
written directly
\begin{equation}
P_{\sf a} = \bra{\Psi} \Proj_{\sf a} \otimes \op\Id \ket{\Psi},
\quad
P_{\sf b} = \bra{\Psi} \op\Id \otimes \Proj_{\sf b} \ket{\Psi}
\label{PaPb}
\end{equation}

Such nonsymmetrical models are even more known due to direct analogue with
Einstein-Podolsky-Rosen consideration \cite{EPR} evolved further in many 
modifications \cite{wiseman} and intensively tested and analyzed till
nowadays \cite{EPRcol}. The arrows on \Fig{depend}b,c
correspond to Einstein's ``spooky action at a distance'' or ``SF mechanism''%
\footnote{Due to Einstein's original term, {\em spukhafte Fernwirkungen}.} \cite{abs}.

\section{Everett's Formulation}
\label{Sec:Ev}

\begin{figure}[htb]
\begin{center}
a)\includegraphics[scale=0.5]{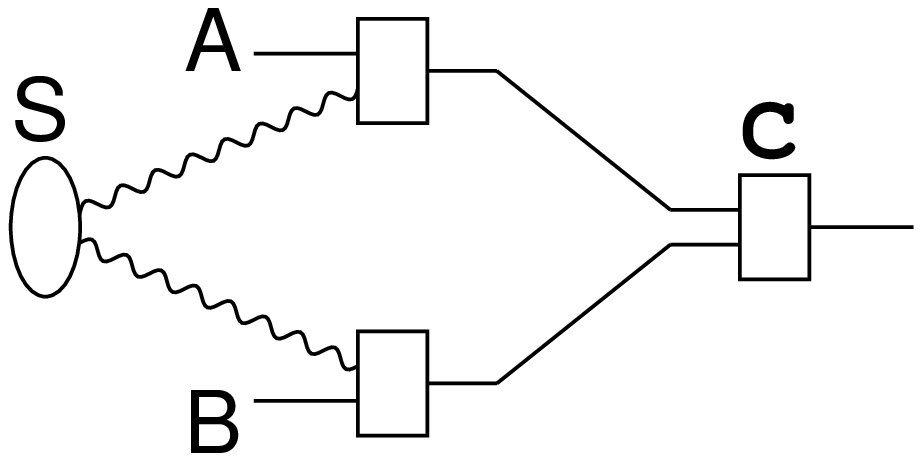}~~ 
b)\includegraphics[scale=0.5]{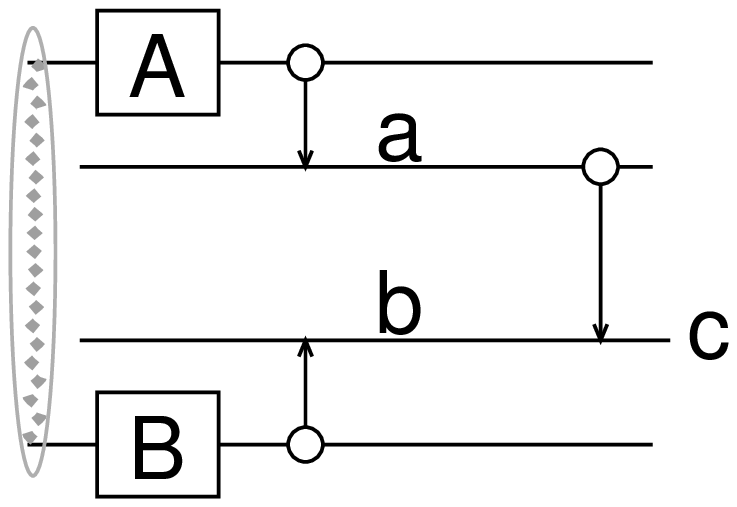} 
\end{center}
\caption{Alternative scheme of measurements}\label{Fig:loccomp}
\end{figure}

A known method to resolve the problem with nonlocality is Everett's formulation 
of quantum mechanics \cite{Ev,DeuHay,Tip0,entrel} and due to difficulties
with standard interpretation, it devotes an accurate consideration.
An essential idea \Fig{loccomp} is a claim, what correlations are
uncovered by some local operation of comparison {\tt C} \Fig{loccomp}a.

A concrete calculations for two spin-half systems may be found in \cite{DeuHay,Tip0,entrel}
with a scheme of {\em quantum network} like \Fig{loccomp}b \cite{DeuHay}.
Analogous model for spin-1 systems used in Conway-Kochen (thought) experiment
is discussed below in App.~\ref{App:CKwor}

In a quantum network model each system denoted as a line (``wire'') and an 
unitary operator as a ``gate'' on one (\pix{-0.25}{0.6}{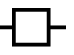}) or 
two (\pix{-0.25}{0.6}{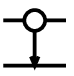}) systems. Such a scheme \Fig{loccomp}b 
illustrates an additional carriers {\sf a} and {\sf b} used for resolution of 
nonlocality problem.

It is not even necessary to {\em completely} accept Everett's formulation. 
Minimal implication of such a model may be suggestion, that it is 
{\em enough} to have {\em entangled outcomes of measurements} 
to explain quantum correlations. 

It may be also raised a question about {\em necessity} of such entanglement. 
Really, all the nonclassical correlations, inequalities and paradoxes may be 
derived from quantum superposition principles. Is it possible to explain the 
correlations between nonlocal outcomes without entanglement, {\em i.e.}, special
kind of superposition?

It is possible to consider a question about treatment of
some quantum correlations as an experimental test of Everett's
interpretation. Yet, such a confirmation of {\em macroscopic superposition}
of different states of measurement devices is rather indirect in comparison 
with other modern experiments providing ``powerful examples for the validity 
of unitary Schr\"odinger dynamics and the superposition principle on increasingly 
large length scales'' \cite{nocol}.

Such indirect evidence could be justified only after proof on impossibility
of other explanations of observed effects. More direct proof should include
description, analysis and testable consequences of explicit models for reduction
process. 

It may appears, that relativity principle prevent any reasonable ``internal''
description of reduction for space-like separated measurements of two observers 
and it is necessary to act in spirit of Copenhagen interpretation and only talk 
about (classical) final results or probabilities. But even lack of possibility to 
introduce some order for such measurements may be overcome in description with 
{\em local clocks}. 

\section{Formalism with Local Clocks}
\label{Sec:twot}

The local clocks, {\em i.e.}, application of two-time formalism to discussion about 
quantum nonlocality is not a new idea, {\em e.g.}, it is represented in two 
last chapters of collection of Bell's papers \cite{BellBook}.

\begin{figure}[htb]
\begin{center}
a)\includegraphics[scale=1.125]{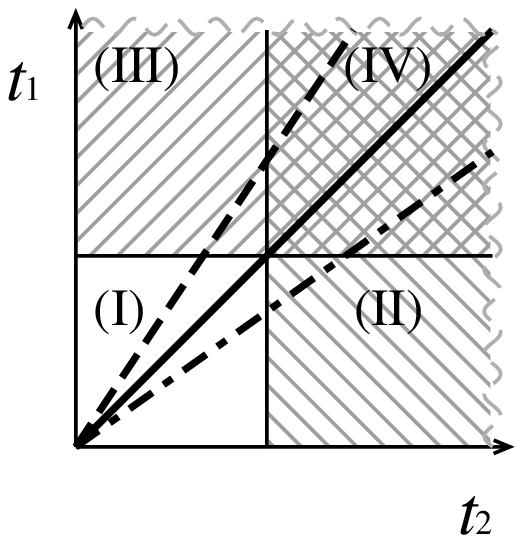}~~
b)\includegraphics[scale=0.75]{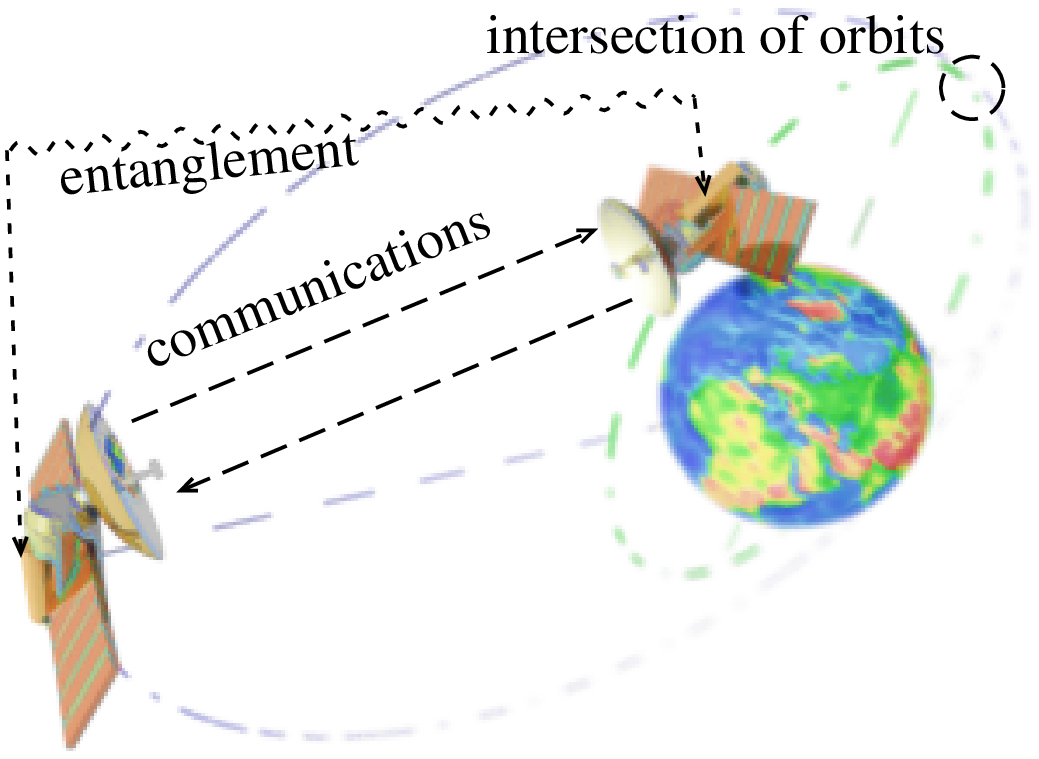}
\end{center}
\caption{Model with local clocks}\label{Fig:twotimes}
\end{figure}

A classical probability (density) for two particles to have 
positions $x_1$ and $x_2$ is represented by a function $\varrho(x_1,x_2)$. 
A dynamical process is usually described via some function $\varrho(x_1,x_2,t)$.
If there is some reason to use separate clocks for each particle,
it is possible to introduce a function $\varrho(x_1,t_1,x_2,t_2)$.

Similar function 
$\varrho(x_1,t_1,x_2,t_2) = \Abs{\psi(x_1,t_1,x_2,t_2)}^2$
may be used in quantum mechanics for description of nonlocality in EPR correlations
\cite[\S21]{BellBook} or GRW models \cite[\S22]{BellBook}. Here the same approach
is used for description of measurement for two parties, if order of event is
not defined. 

Let $\op{\rho}(x_1,t_1,x_2,t_2)$ is a density matrix for such a system. 
It is suggested, that measurement is performed by each party on time $\tau$ on local
clock. So there are four zones on ``two-time plane'' \Fig{twotimes}a: 
(I) $t_1 < \tau,\ t_2 < \tau$, (II) $t_1 < \tau,\ t_2 \ge \tau$,
(III) $t_1 \ge \tau,\ t_2 < \tau$, (IV) $t_1 \ge \tau,\ t_2 \ge \tau$.

Quite realistic example of such a scheme is entangled state used in 
experiments with satellites \Fig{twotimes}b. Difference between clocks due 
to relativistic effects in principle may be essential in modern experiments.
Three lines (solid, dashed, dash-and-dot) on \Fig{twotimes}a represent 
different method of synchronization of $t_1$ and $t_2$ with some global 
clock $t$. 

Equipment on both satellites may be programmed to performs 
measurement at the same time $\tau$ on local clocks, but order is not
an absolute property due to relativity principle. Quantum cryptography with
entangled states and free space communications \cite{FS144} may make such
schemes quite actual. Even if pair of entangled qubits used in such 
communications is not relevant with particular scheme suggested in \cite{CK}, 
systems of quantum cryptography with qutrits \cite{qu3en,qu3cr} are already
valid area for such a discussion.

Let us denote measurement procedures as ${\tt M}_\A$ and ${\tt M}_\B$ for first
and second party respectively. If initially there is some density matrix $\op{\rho}$
and it is changed only due to measurements, then for each zone it may be written
$\op\rho_{I}=\op\rho$, $\op\rho_{II} = {\tt M}_\B \op\rho$, 
$\op\rho_{III} = {\tt M}_\A \op\rho$, and for independent measurements 
$\op\rho_{IV} = {\tt M}_\A {\tt M}_\B \op\rho = {\tt M}_\B {\tt M}_\A \op\rho$. 

\section{Von Neumann Measurements}
\label{Sec:vnNm}

The measurements above are presented in rather abstract way. It is possible
first to consider von Neumann measurement \cite{vnNeu} defined for
single system and a measurement basis $\varphi_k$ as a linear map
\begin{equation}
 \op{\rho} \mapsto 
 \sum_k \bra{\varphi_k} \op{\rho} \ket{\varphi_k} \Proj_{[\varphi_k]},
 \quad \Proj_{[\varphi_k]} \equiv \ket{\varphi_k}\bra{\varphi_k}.
\label{vnNeuMes}
\end{equation}

For more general case of incomplete von Neumann measurement \cite{wer}
it may be rewritten as
\begin{equation}
\op{\rho} \mapsto \SPr(\op\rho) = \sum_m \SPr_m(\op\rho), \quad
\SPr_m(\op\rho) = \Proj_m \op\rho\,\Proj_m,
\label{incoMes}
\end{equation}
where $\Proj_m$ is not necessary one-dimensional projector. Now
it is possible to define measurements for two-parties system\footnote{
Such definition is possible due to properties of tensor product, because 
$\SPr$ is {\em linear} transformation of operators (sometimes 
it is called ``superoperator'').} 
\begin{equation}
 {\tt M}_\A = \SPr_\A \otimes \sop\Id, \quad
 {\tt M}_\B = \sop\Id \otimes \SPr_\B,
\label{SMAMB}
\end{equation}
where $\sop\Id = \op\Id$ is identity (``unit superoperator''),
$\SPr_\A$ and $\SPr_\B$ are defined by \Eq{incoMes} using set of projectors 
for measurements of $\A$ and $\B$.

Let us consider an example with two possible decompositions of the 
same entangled state
\begin{equation}
 \ket{\Psi} = \sum_k \alpha_k \ket{a_k} \ket{u_k} = \sum_j \beta_j \ket{v_j} \ket{b_j},
\label{twosums}
\end{equation}
there $\ket{a_k}$ and $\ket{b_j}$ are measurement bases chosen by $\A$ and $\B$
respectively. It is now possible to find values of density matrixes in all
four zones: 
\begin{subequations}
\renewcommand{\theequation}{\theparentequation$_{\Roman{equation}}$}
\begin{gather}
\op\rho_{I} = \op\rho_{[\Psi]} = \ket{\Psi}\bra{\Psi},\\ \textstyle
\op\rho_{II} = \sum_j |\alpha_j|^2\,\ket{v_j}\bra{v_j} \otimes \ket{b_j}\bra{b_j},
\label{rzII}\\ \textstyle
\op\rho_{III} = \sum_k |\beta_k|^2\, \ket{a_k}\bra{a_k} \otimes \ket{u_k}\bra{u_k},
\label{rzIII}\\ \textstyle
\op\rho_{IV} = \sum_{k,j} q_{kj}\,\ket{a_k}\bra{a_k} \otimes \ket{b_j}\bra{b_j},
\label{rzIV}
\end{gather}
\label{rz}
\end{subequations}
where $q_{kj}$ in \Eq{rzIV} is some matrix. It may be expressed in two
different ways
\begin{equation}
q_{kj} = \Abs{\alpha_j\brkt{a_k}{v_j}}^2
= \Abs{\beta_k\brkt{u_k}{b_j}}^2
\label{qkj}
\end{equation}
using \Eq{rzII} and \Eq{rzIII} respectively.

It is possible to use particular symmetric case of \Eq{twosums} 
\begin{equation}
 \ket{\Upsilon} = \frac{1}{\sqrt{N}}\sum_{k=1}^N \ket{a_k} \ket{a_k} 
  = \frac{1}{\sqrt{N}}\sum_{k=1}^N \ket{b_k} \ket{b_k},
\label{symsums}
\end{equation}
appropriate for discussion on Conway-Kochen theorems \cite{CK} (see \Eq{Stwin} and
\Eq{Twinv} in App.~\ref{App:CKwor}). In such a case \Eq{rz} become simpler
\begin{subequations}
\renewcommand{\theequation}{\theparentequation$_{\Roman{equation}}$}
\begin{gather}
\op\rho_{I} = \op\rho_{[\Upsilon]}=\ket{\Upsilon}\bra{\Upsilon},\\ \textstyle
\op\rho_{II} = N^{-1}\sum_{j=1}^N \ket{b_j}\bra{b_j} \otimes \ket{b_j}\bra{b_j},
\label{rsII}\\ \textstyle
\op\rho_{III} = N^{-1}\sum_{k=1}^N  \ket{a_k}\bra{a_k} \otimes \ket{a_k}\bra{a_k},
\label{rsIII}\\ \textstyle
\op\rho_{IV} = N^{-1}\sum_{k,j=1}^N \Abs{\brkt{a_k}{b_j}}^2
              \ket{a_k}\bra{a_k} \otimes \ket{b_j}\bra{b_j},
\label{rsIV}
\end{gather}
\label{rsym}
\end{subequations}

Scheme of measurement \Fig{twotimes}a may be represented
\begin{equation}
 \begin{array}{ccc|cc}
  {}^{(III)} &
   \MesPar{a}{a} &\longrightarrow& 
  {}^{(IV)} &
   \MesPar{a}{b} \\ \hline
  &
  \uparrow & \nearrow & &\uparrow \\ 
  {}_{(I)} &
  \op\rho_{[\Upsilon]} &\longrightarrow& 
  {}_{(II)} &
   \MesPar{b}{b}\\
\end{array}
\label{mesI-IV}
\end{equation}
where $\MesPar{\cdot}{\cdot}$ is brief notation for sums in \Eq{rsym}
and also may be considered in more general sense as an abstract representation
of result of measurement(s). 

\section{Definite Outcomes}
\label{Sec:defout}

Introduction of von Neumann measurement only partially illustrates problem of
reduction. It is used description of ensembles and transition between quantum
mechanical description and classical statistical properties are given by
known equations, like Born rule. Nonlocality of such equations is often
demonstrated only indirectly and related with additional assumptions
about class of reasonable local models \cite{BellEPR,BellHidd,CHSH}. 

Formally, description with ensembles still contains some analogue of multiple 
world formalism, because it is yet some equation with superposition of different states.
The problem becomes more difficult, if to try to find a method for generation
of certain outcome for {\em each measurement}. It is called sometimes
{\em the problem of definite outcomes} \cite{Schl}.

In quantum mechanical description without reduction such a question looks 
incorrect, because there is no certain outcome in superposition. In description
with ensembles such question may be correct, but often is weaved as something 
inappropriate. It is analogue of classical statistical theory with description
of probabilities, but not reasons for given outcome in single test.

More abstract measure theory may represent an alternative view, then distribution
of probabilities decribes all possible outcomes with different measure of
``actuality.'' Bell represents usual argument against such a view \cite[\S11]{BellBook}
``Whereas Everett assumes that all configurations of his special 
variables are realized at any time, each in the appropriate branch universe, 
the de Broglie world has a particular configuration. 
I do not myself see that 
anything useful is achieved by the assumed existence of the other branches 
of which I am not aware.''  

In cited paper Bell just discussed de Broglie-Bohm theories, {\em i.e.}, 
models with possibility to assign certain outcome to each measurement.%
\footnote{Yet, it uses some nonlocality \cite{wiseman}.}  Certainly,
usual scientific principle of ``Occam's razor'' demands to ``cut'' all
extra branches. On the other hand, we may be not aware about some phenomena, 
but have to accept that as most reasonable way to explain observed things. 

\medskip
  
The problem with suggestion about only one term in superposition may be
illustrated with Conway-Kochen experiment \cite{CK}. It may be used
particular case of \Eq{symsums} with three terms. 
\begin{equation}
\Upsilon = \frac{1}{\sqrt{3}}\bigl(\ket{1}\ket{1}+\ket{2}\ket{2}+\ket{3}\ket{3}\bigr)
\label{Ups}
\end{equation}
See App.~\ref{App:CKEx} and App.~\ref{App:CKwor} and Ref.~\cite{Sim} for 
more details. 

First observer uses for measurement three basic states $\ket{x}$,
$\ket{y}$ and $\ket{z}$ simply associated with three orthogonal axes and
second observer is using $\ket{x'}$, $\ket{y'}$, and $\ket{z'}$. Let's 
consider particular situation with $\ket{x'}= \ket{y}$, 
$\ket{y'} = (\ket{x}+\ket{z})/\sqrt{2}$, $\ket{z'} = (\ket{x}-\ket{z})/\sqrt{2}$.

Final (in ``zone IV'') distribution of probabilities \Eq{rsIV} may be 
represented by table 
\begin{equation}
   \frac{1}{3}\left(
   \begin{array}{ccc}
     \Abs{\brkt{x}{x'}}^2 & \Abs{\brkt{x}{y'}}^2 & \Abs{\brkt{x}{z'}}^2 \\
     \Abs{\brkt{y}{x'}}^2 & \Abs{\brkt{y}{y'}}^2 & \Abs{\brkt{y}{z'}}^2 \\
     \Abs{\brkt{z}{x'}}^2 & \Abs{\brkt{z}{y'}}^2 & \Abs{\brkt{z}{z'}}^2 \\
  \end{array}
  \right)  
  =
   \frac{1}{3}\left(
   \begin{array}{ccc}
     0 & 0.5 & 0.5 \\
     1 & 0 & 0 \\
     0 & 0.5 & 0.5 \\
  \end{array}
  \right)
\label{tab3x3}
\end{equation}

In table \Eq{tab3x3} some combinations of outcomes have zero probabilities
in zone (IV), {\em e.g.}, $\MesPar{x}{x'}$. On the other hand, impossibility
to have outcome $x$ for first observer is certain only for particular choice
of the second one. In zone (III) outcomes $x$, $y$ and $z$ have equal
weights, but may be impossible in combination with certain choice and outcome
of second observer. The analogue is true for second observer, zone (II). 

If to keep all terms in superposition, it could be considered as some
formal redistribution, like $(1/3,1/3,1/3) \mapsto (1,0,0)$ or
$(1/3,1/3,1/3) \mapsto (0,1/2,1/2)$,  
where combinations like $\MesPar{x}{x'}$ are excluded from final mixture at all. 
With definite outcome for each observer it is not clear,
why the {\em impossible pairs} like $\MesPar{x}{x'}$ could not be produced,
if both $x$ and $x'$ are {\em valid} outcomes.

\section{Conway-Kochen Theorems}
\label{Sec:CKT}

Example below raises question about possibility to test existence  
of ``invisible terms'' {\em indirectly}. Such a question lay beyond the limits 
of quantum theory, there is not defined a receipt to choose
and save only single term in some superposition. It may be 
discussed in some extensions like de Broglie-Bohm theories 
already mentioned above in Bell's cite or Ghirardi-Rimini-Weber (GRW) 
models \cite{DynRed}.

In \cite{CK,CK2v2,CK3} problem of nonlocal measurements was
analysed also with relation to GRW theories and it was shown, that for wide
class of theories (``Free State theories'') it is impossible to
suggest {\em local} method to provide certain outcomes for pair
of measurements of some entangled state in agreement with 
SPIN and TWIN axioms, considered as simple consequence 
of quantum theory.

It may be shown, that both SPIN and TWIN axioms
formally do not respect Lorentz invariance \cite{Wig}, and so it is better to
use limit of small speeds and formulation without appeal
for relativity, like \cite{CK3}. After all, only field theory may
properly take into account relativistic effects \cite{Wei1} and so
reasoning about isolated pair of particles may have some flaws.
Yet another criticism may be found in \cite{Rep,Rep2}.

In fact, problem with ``reconciliation'' of quantum mechanics and 
special relativity is well known. One subject of Nobel Prize Lecture 
{\em Asymptotic freedom: From paradox to paradigm} \cite{Wilczek} by 
Frank Wilczek in 2004  had quite illustrative title: 
{\em ``Paradox 2: Special Relativity and Quantum Mechanics Both Work''}
and declared ``creative tension'' of such paradoxes, leading to four Nobel
Prizes (1933, 1965, 1999, 2004).

So Conway-Kochen arguments illustrating some problems with 
nonlocality even in nonrelativistic quantum mechanics \cite{CK3}
devote accurate consideration, especially because it is not suggested yet, 
how the field theory could avoid that.

A consequence of Conway-Kochen theorem necessary for present purpose is 
absence of ``reasonable'' local model with definite outcomes for state 
\Eq{Ups} with three terms.
In \cite{CK,CK2v2,CK3} instead of term ``reasonable'' was used idea
of ``free state theories'' describing ``the evolution of a state 
from an initial arbitrary or ``free'' state according to laws that are themselves 
independent of space and time'' \cite{CK}.
In App.~\ref{App:CKEx} is reproduced some arguments to do not limit problem
in such a way and to use ``any'' instead of ``reasonable'' above.  

Anyway, experiment with two entangled spin-1 particles suggested by Conway and 
Kochen \cite{CK,CK2v2,CK3} is very hard to explain by some analogue of 
de Broglie-Bohm theory and so argument of Bell {\em et al} against Everett 
multiple world formalism is not strong enough, because models with single 
branch may look much more weird than multiverse ``science fiction.''

Authors of \cite{CK,CK2v2,CK3} are embarked in discussion about free
will of particles and experimenters, but it is possible to suggest analogues
of they theorems without any relation with possibility of free choice. It is
also possible to argue, what accepting ``free will'' for particles in given
experimental setup would produce possibility of superluminal information 
exchange, because it is suggested some coordinated choice between pair of 
outcomes like (0,0) or (1,1).

It is clear from analysis of \cite{CK,CK2v2,CK3} that even using instead 
of particles two intelligent human gamblers (Carl and Clara) with free 
will and perfect knowledge of quantum mechanics could not establish 
necessary correlations (if to exclude cooperation with Alice and Bob 
like preliminary agreement about setup of all future measurements). 
So term ``free will'' in \cite{CK,CK2v2,CK3} is rather used for some 
mysterious and subtle ability to make right choice without necessary 
information.

Here Conway-Kochen result is discussed due to very clear demonstration
of contradiction between nonlocality and correlation of {\em definite outcomes}.
In \cite{CK,CK2v2,CK3} a possibility of macroscopic quantum superposition 
between results of measurements is even not considered, but it is valid
explanation of necessary correlations compatible with quantum mechanics. 

A bias against Everett interpretation is so strong, that as more
appropriate explanation are considered almost supernatural 
properties of elementary particles \cite{CK,CK2v2,CK3}. 
In \cite{april} is suggested more ``precise'' statement of related idea: 
``quantum randomness can be controlled by influences from outside spacetime, 
and therefore by immaterial free will.''

The purpose of given paper is not an apology for multiverse. The idea of 
multiverse looks like some try of classical reasoning about quantum 
superposition in single universe. The more precise statement of problem 
is possibility to use macroscopic superposition of measurement outcomes
as an explanation of quantum correlations observed in experiments.

Such macroscopic superposition may not be accepted without doubts. 
The possible loopholes often discussed
in relation with local hidden variables models \cite{Genov}. 
As an support for possibility of future achievements in next section is 
reproduced yet another ``extravagant'' local model.

It is astonishingly yet, that problem with theoretical model
of reduction and definite outcomes may convert interpretation of
many experiments claiming confirmation of quantum correlations with
too high precision into argument against ``standard'' interpretation%
\footnote{Standard or ``orthodox'' interpretation is usually formulated with 
wave function ``collapse'' during measurements and Copenhagen variant also 
suggests ``necessity of classical concepts'' \cite[{\S}IV.B]{Schl}. It should
be mentioned yet a ``black-box'' version of standard interpretation often
used by ``practicing physicists'': there are very precise and correct formulas for
calculation probabilities for outcomes of quantum processes and discussion
about interpretations is time consuming curiosity without hope on constructive
conclusions and practical applications. The cites about probabilities and
amplitudes from popular Feynman lectures are very
illustrative \cite{FeyQED}: ``Does this mean that physics, a science of great 
exactitude, has been reduced to calculating only the {\em probability} of an event, 
and not predicting exactly what will happen? Yes. That's a retreat, 
but that's the way it is: Nature permits us to calcalculate only probabilities.'',
\dots, ``Furthermore, all the new particles and new 
phenomena that we are able to observe fit perfectly with 
everything that can be deduced from such a framework of 
amplitudes, in which the probability of an event is the 
square of a final arrow whose length is determined by  
combining arrows in funny ways (with interferences, and so 
on). So this framework of amplitudes has {\em no experimental 
doubt} about it: you can have all the philosophical worries 
you want as to what the amplitudes mean (if, indeed, they 
mean anything at all), but because physics is an  
experimental science and the framework agrees with experiment, 
it's good enough for us so far.''} 
and an additional evidence for Everett's formulation (if other alternatives 
may be even more strange). It may be considered as an argument to revisit 
results of even standard experiments with more attention \cite{aff}.

\section{Extended Probabilities}
\label{Sec:negprob}

Quantum state of composite system is called {\em separble} \cite{wer} if density 
matrix may be represented as sum
\begin{equation}
 \rho = \sum_k \lambda_k \rho_k^\A \otimes \rho_k^\B, \quad \lambda_k > 0.
\label{sep}
\end{equation}
For such a state nonlocal measurement discussed above may be modeled by local
model, if to suggest that source emits different
pairs of states $\rho_k^\A \otimes \rho_k^\B$ with probabilities $\lambda_k$. 

On the other hand, if to exclude requirement about positivity of $\lambda_k$ 
in \Eq{sep} any state may be represented as \cite{rob,VlPr}
\begin{equation}
 \rho 
  = \sum_{k=1}^{n^+} \lambda^+_k \op\rho_k^\A \otimes \op\rho_k^\B
 - \sum_{k=n^+ + 1}^{n^+ + n^-} |\lambda^-_k|\, \op\rho_k^\A \otimes \op\rho_k^\B
 = \kappa\op\rho^+ - (\kappa-1)\op\rho^-, 
 \quad \lambda^+_k > 0,
 \quad \kappa=\sum_{k=1}^{n^+}\lambda^+_k
\label{insep}
\end{equation}
with $\op\rho^+$ and  $\op\rho^-$ are two separable states. In App.~\ref{App:Decomp}
is shown such representation of state \Eq{Ups} for $N=3$ with 
$n^+=18$, $n^-=15$, and $\kappa=7$.

Association of entangled states with negative probability is not a new
thing. Talk of Feynman about such negativity for Bell state \cite{FeySim} 
on PhysComp'81 may be included into origins of modern quantum information 
science and quantum cryptography. It was also discussed by Bell in 
a paper about two-time Wigner distribution and EPR already 
mentioned above \cite[\S21]{BellBook}.

\begin{figure}[htb]
\begin{center}
\includegraphics{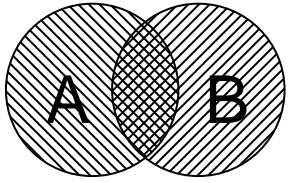} 
\end{center}
\caption{Venn diagram for $\A$, $\B$, $\A \cup \B$, $\A \cap \B$}\label{Fig:AandB}
\end{figure}

In later paper \cite{FeyNeg} Feynman mentioned more examples in classical and
quantum mechanics to conclude ``\ldots conditional probabilities and probabilities
of imagined intermediary states may be negative in a calculation of probabilities
of physical events and states.'' 
Mathematical justification for such formalism is theory
of {\em signed measures (charges)} \cite{KolmFom}.
Another trivial example is usual classical 
equation for calculation of probability for union of two overlapped classes 
of events $P(\A \cup \B) = P(\A) + P(\B) - P(\A \cap \B)$ \cite{Fell} 
with negative sign before probability of events in class $\A \cap \B$,
see \Fig{AandB}.

Table \Eq{Decomp81} in App.~\ref{App:Decomp}
shows signed decomposition \Eq{insep} for a state $\ket{\Upsilon}$ 
\Eq{Ups} used in examples above and related with Conway-Kochen theorems. For 
given state and choice of basis in space of density matrixes there are 18 positive 
and 15 negative terms. Formally, it is possible to use a local theory for calculation 
of some averages for physical values in agreement with quantum mechanics. Really, it
is possible to represent
\begin{equation}
  \op\rho_{[\Upsilon]} \equiv \ket{\Upsilon}\bra{\Upsilon} = 
  7\, \op\rho^+_{[\Upsilon]} - 6\, \op\rho^-_{[\Upsilon]}.
\label{rhoUps}
\end{equation}
with some separable, {\em i.e.}, {\em classically correlated} states $\rho^\pm_{[\Upsilon]}$.
So there is local hidden variable model for both states. It just two sources, emitting
pairs states $\op\rho_k^\A \otimes \op\rho_k^\B$ with probabilities of $\lambda^+_k/\kappa$ and
$|\lambda^-_k|/(\kappa-1)$ respectively. Now if average of some operator $\op{O}$
for each source $\Av{\op O}_{[\Upsilon]}^\pm = \Tr(\op\rho^\pm_{[\Upsilon]}\op{O})$, 
due to \Eq{rhoUps} it is possible to write 
\begin{equation}
\Av{\op O}_{[\Upsilon]} = \bra{\Upsilon}\op O\ket{\Upsilon} = 
7\, \Av{\op O}_{[\Upsilon]}^+ - 6\, \Av{\op O}_{[\Upsilon]}^-. 
\label{AvOpm}
\end{equation}

Such two sources also could be joined in single one with two kinds of events: 
first kind with probabilities $\lambda^+_k$ increases some counter, second 
one with probabilities $|\lambda^-_k|$ decreases that. 
Some other ideas about work with negative probabilities my be also found
in already mentioned Feynman publications \cite{FeySim,FeyNeg} and
recent Hartle's work on extended probabilities \cite{ExtProb}.

Sometimes such models may guarantee correct positive probabilities for 
observable events {\em only} for big number of trials and it is certain problem. 
Let us consider measurement of state \Eq{Ups} discussed above. If
both parties use the same basis, probabilities of possible outcomes
are represented as $P(k,k)=1/3$, $P(j,k)=0$, $j \neq k$, $j,k=1,\ldots,3$. 
It is also defined probability for each party for given outcome 
$P_\A(k) = P_\B(k) \equiv P(k) = 1/3$.

Table \Eq{Decomp81} used for local modeling contains 33 different
types of events: 18 for increasing counter and 15 for decreasing one.
Mathematical expectation values for number of events with increasing
and decreasing counters are $\Av{N^+}=\kappa = 7$ and 
$\Av{N^-}=\kappa-1 = 6$ respectively. It is also possible to 
calculate values for number of events for any choice
of measurement parameters. 

For the measurement basis $\ket{j}\ket{k}$ it may be found
$\Av{N^+(k,k)}=4/3$, $\Av{N^-(k,k)}=1$, $\Av{N^\pm(j,k)}=1/2$,
$j \neq k$, $j,k=1,\ldots,3$. Expectations for each party are
also may be defined $\Av{N^+(k)}=7/3$, $\Av{N^-(k)}=2$.
It produces correct probabilities
$P(j,k)=\Av{N(j,k)}=\Av{N^+(j,k)} - \Av{N^-(j,k)}$,
$P(k)=\Av{N(k)}=\Av{N^+(k)} - \Av{N^-(k)}$ for big number 
of trials, but for single experiment it may not guarantee a proper behavior.

In usual experiment we should expect appearance of only one event
between three possible combinations (1,1), (2,2), (3,3). In ``balanced''
model we have integer $N(j,k)=N^+(j,k) - N^-(j,k)$ and it even may be 
negative number. It is only in average $P(j,k)=\Av{N(j,k)}$, but incorrent 
numbers may be generated in single trial, {\em i.e.},  $N(j,k) \ne 0$ for $j \neq k$,
$N(1,1)+N(2,2)+N(3,3) \neq 1$ and, worst, $N(j,k) < 0$.

Even if problem with negative counters could be resolved using additional 
correlation between ``positive'' and ``negative'' events, impossibility 
of completely correct simulation of measurement may be consequence of 
Conway-Kochen theorem \cite{CK}. On the other hand, if improved model 
would generate nonnegative numbers, but total number of events 
$N = N^+ - N^-$ is ``wrong'' $N \ne 1$ (but $N \ge 0$), it could be 
considered as realistic situation with registration of more than one 
particle ($N > 1$) or failure to detect something ($N=0$). For such a 
stochastic case arguments of Conway-Kochen theorem already may not be 
applied.

\section{Quantum Nonlocality and Inseparability}
\label{Sec:insep}

Despite of use ``nonlocality'' term, in formulation of most problems discussed
here was not even used any coordinates. More certain reason for some problems 
is lack of clear analogue of classical conception of {\em subsystem} in
quantum mechanics.

Such quantum inseparability related with definition of compound system 
in quantum mechanics using tensor product. Contradictory properties of such 
construction are remarkably exploited in famous EPR paper \cite{EPR}. 

It is strange to read in some modern papers 
interpretation of EPR view about quantum states as ``narrow-minded'' \cite{brinc},
because discussed problem is rather direct consequence of some mathematical 
structures used in quantum mechanics. 

Formally, in theory of categories \cite{slang} tensor product {\em is 
not product} and so unlike direct product in classical physics it produces 
a problem with discussion about ``parts of bipartite system.'' 
Using some manipulations with dual spaces, {\em i.e.}, combining ``bras'' and ``kets,'' 
it is possible only partially resolve some
problems for Heisenberg representation or density matrix formalism. 
For each part we may associate operators with elements of some subspaces 
using notations like $\op A \otimes \op\Id$ and $\op\Id \otimes \op B$
and introduce spaces of operators for each subsystem.

However, operation of partial trace for preparation of reduced density matrix
of subsystem may produce a useless result, {\em e.g.}, both for Bell
state and Conway-Kochen-Peres state $\ket{\Upsilon}$ \Eq{Ups},
it produces maximally mixed state proportional to unit, but 
it is an equivalent of a system in an ``unknown'' state.

More technical manifestation of problem with tensor product is
exponentially big number of parameters for compound system:
union of two classical systems with dimensions $N_1$ and $N_2$ 
may be described by $N_1 + N_2$ parameters,
but dimension of tensor product for two spaces is $N_1 N_2$.\footnote{Formally, 
tensor product are also indirectly used in classical physics in 
description of probability densities $\rho(x_1,x_2)$ for two systems
already mentioned earlier, but it is only secondary construction, defined 
as space of functions on well-defined direct product with set of pairs $(x_1,x_2)$.} 

For pair of spin-1 particles $N_1=N_2=3$ such consideration even more
actual, than for qubits ($N_1=N_2=2$).
If such bipartite system is local, we could not discuss an internal structure
and consider some indivisible object with nine-dimensional Hilbert space,
but for two separated systems without interaction problem with ``superfluous''
dimensions is more visual. 
 
Nonlocality in such a consideration is only tool to emphasize the problem
with separability. It is not clear, if acceptance of nonlocal theories
could resolve all paradoxes. Let's suggest existence of some nonlocal model
for spacelike separation. It must take into account choice of both observers
for generation of outcomes. At least one measurement for such a model should
take into account choice of other party. It is possible to denote this party  
for certainty as $\A$.

Let us consider some analogue of twin paradox. Two satellites \Fig{twotimes}b 
at the same moment on local clocks are landing and performing measurements. 
Let for local clock on the Earth satellites $\A$ and $\B$ are
landing to the same position at $t_\B > t_\A$. It is timelike separation,
but if our model should resolve problem with inseparability without suggestion
about kind of spacetime interval between events, we should accept dependency 
of measurement $\A$ form choice of $\B$ declared above. In such a case 
measurement $\A$ should depend on some future events.

Otherwise we could suggest, that equipment of each satellite should resolve
if measurements are spacelike separated or not. Such model may include
some unknown particles emitted at moment of measurements and travelling
with speed of light. It is a model irrelevant with discussion above,
because it may not resolve problem with spacelike separated measurements.

Finally, we should conclude that try to use nonlocality for resolution of
spacelike separated measurement due to relativistic effects may result 
appearance of paradoxes with action backward in time.

\section{Conclusion}
\label{Sec:concl}

It may be appropriate here a citation in relation with some topics 
of quantum field theory (the claster decomposition principle) \cite[\S4.3]{Wei1}: 
``It is one of the fundamental principles of physics (indeed, of all science) 
that experiments that are sufficiently separated in space have unrelated 
results. The probabilities for various collisions measured at Fermilab 
should not depend on what sort of experiments are being done at CERN 
at the same time. If this principle were not valid, then we could never 
make any predictions about any experiment without knowing everything 
about the universe.'' 

These words were written in a handbook on quantum theory not so long 
time ago, but experiments with photons separated over 144 km just display 
the specific quantum correlations for sufficiently separated measurements
\cite{FS144}. Current technology may even test some preliminary ideas about
influence of `correct quantum gravity' effects on state vector reduction
\cite{ENM} for spacelike separated correlation on distance 18 km resulting 
``displacement of a macroscopic mass'' \cite{GrCol}.

Resolution of problems with such correlations and description of phenomena 
``without knowing everything about the universe'' (possibly, including 
future events) is not quite obvious. It is concluded in \cite{wiseman} that
Bell's theorem ``forces us to choose between a nonlocal interpretation of QM 
(either well-defined like the Bohmian interpretation, or ill-defined like 
Bohr's Copenhagen interpretation), and extreme subjectivism.''
The Conway-Kochen theorem made problem
with nonlocality in Copenhagen interpretation even more clear, if it 
``prevents the existence of local mechanisms for reduction'' \cite{CK}.

Anyway, the work also demonstrates an evidence
for an optimistic view about possibility to overcome some problems. 
It was mentioned in Sec.~\ref{Sec:Nonloc} that quite simple nonlocal
model exists for any quantum system. 
In Sec.~\ref{Sec:Ev} was recollected local model without collapse 
originated by Everett formulation of quantum mechanics. It is
discussed with more details in App.~\ref{App:CKwor} for 
Conway-Kochen scheme with two entangled spin-1 particles.
In Sec.~\ref{Sec:twot} is revisited nonlocal description of 
collapse without fixed order of events. So incompatibility 
between relativity principle and reduction process 
sometimes may be overestimated. Finally, extended stochastic models 
discussed in Sec.~\ref{Sec:negprob} demonstrate possibility of 
achievements in rather unexpected directions.

\newpart
\appendix
\section*{Appendixes}

\section{Analysis of Conway-Kochen Scheme}
\label{App:CKEx}

A thought experiment considered in \cite{CK} let us raise problem with quantum
nonlocality quite sharp: {\em it is impossible to suggest a local model with
definite outcomes of measurements for two entangled particles with spin one}. 

\subsection{Conway-Kochen Theorem Revisited}

This serious implication may be derived with axioms suggested in 
\cite{CK,CK2v2,CK3}. In this experiment two parties (\A\ and \B, {\em e.g.}, Alice and Bob) 
are measuring square of spin in different directions. It may be zero or one. 
Let's denote that as $J^2_v{}_\A$ and $J^2_w{}_\B$ for measurements in direction 
$v$ for \A\ and $w$ for \B\ respectively.

{\em A local model} for such experiments should describe independent methods 
to assign values of $J^2_v{}_\A$ and $J^2_w{}_\B$ for {\em any} $v$ and $w$ and, 
so, it is defined two {\em functions} $\JJ_n^\A(v)$ and $\JJ_n^\B(w)$. 
Here $n$ is number of experiment in a series, because measurement of the same 
direction in different experiments {\em should not} necessary produce the same value. 
   
Due to the TWIN axiom for the same direction  two measurements should
produce the same result: 
$J^2_v{}_\A = J^2_v{}_\B.$
Such condition may be satisfied if and only if local model uses two equal functions 
$\JJ_n^\A = \JJ_n^\B \equiv \JJ_n$, but experiment may be more complicated. 

For first system it is possible at the same time to measure spins in three orthogonal 
directions $v_1$, $v_2$, $v_3$ and TWIN axiom is valid for any $v_k=w$, $k=1,2,3$.
So for three orthogonal directions local model should use the same function 
$\JJ_n(v_k)$. On the other hand, measurements of squares of spins in three orthogonal
directions due to SPIN axiom always should produce values 1,0,1 in some order and,
so, functions $\JJ_n(v)$ should have the same property.

Due to Kochen-Specker theorem \cite{KS} function with such property does not exist,
{\em i.e.}, it is impossible to assign values \{0,1\} to a sphere to guarantee 
two units and one zero associated in some order with any triple of orthogonal 
vectors. 

Some basic ideas of Conway-Kochen theorem(s) \cite{CK,CK2v2,CK3} are recollected
above for completeness and will be used further (but not ``FREE'' conclusion). 
To avoid some difficulties in discussion
about relativity, causality, information transfer, {\em etc.}, initial FIN axiom
about finite speed of information transfer \cite{CK} was changed further to 
MIN axiom about impossibility of mutual influence of measurements performed
by \A\ and \B\ \cite{CK3}.

The MIN axiom is simply axiom of locality, postulated without additional
explanations about reason of such locality.
The difference of nonlocal models is possibility to use instead of
$\JJ_n^\A(v)$ and $\JJ_n^\B(w)$ pairs like $\JJ_n^\A(v_\A;w_\B)$ and $\JJ_n^\B(w_\B;v_\A)$.
For such a symmetric case (\Fig{depend}a) pair of results of measurements 
may be modeled by a stochastic function: 
$(v_\A,w_\B) \mapsto (J^2_v{}_\A, J^2_w{}_\B)$ \cite{Sim}. 

Models with nonsymmetrical pairs like $\JJ_n^\A(v_\A;w_\B)$ and $\JJ_n^\B(w_\B)$ 
(\Fig{depend}b) or $\JJ_n^\A(v_\A)$ and $\JJ_n^\B(w_\B;v_\A)$ (\Fig{depend}c) are 
also possible. They may be described using equations for 
conditional probabilities \Eq{condprob} and ``SF mechanism'' 
discussed earlier.
 
Authors of the Free Will theorem \cite{CK} suggest to avoid nonlocality using
idea about some ``spontaneous'' information. If to denote such 
``spontaneous variables'' as $\mathfrak{f}_a$, $\mathfrak{f}_b$, it may be defined a pair 
of functions $\JJ_n^\A(v_\A;\mathfrak{f}_a)$ and $\JJ_n^\B(w_\B;\mathfrak{f}_b)$, 
but it is not clear, if such situation may be distinguished from nonlocal model 
with pair $\JJ_n^\A(v_\A;w_\B)$ and $\JJ_n^\B(w_\B)$, discussed earlier. 

\subsection{Free State Theories, Functions and Relations}

The Free State theorem \cite{CK} provides more explicit claim about {\em impossibility} 
to use ``free state,'' {\em i.e.}, ``functional'' or ``evolutional'' approach to description 
of considered experiments with two spin-1 particles. It may be formulated with notation
above as impossibility to suggest local functions $\JJ_n^\A(v)$ and $\JJ_n^\B(w)$.

The problem with Free State theorem is negative formulation of main statement: there are
no free state theories with necessary properties. It {\em does not} mean, what some
``non-free-state'' theory, {\em e.g.}, models with ``free will,'' may satisfy suggested 
axioms. Let's discuss this problem briefly.

Janus model, used in \cite{CK} is interesting for discussion about hidden
violation of Lorentz invariance and FIN principle, but directly contradict 
to MIN axiom in \cite{CK3} and so may not be used for proof of consistency.
It is necessary to consider more general arguments and analogues
of Janus model.

Simplest generalisation of functional dependence is {\em relation}, {\em i.e.}, 
set of pairs $(x,y) \in R$. The function is particular case of relation
with all first elements are different. A function $y = f(x)$ may be written as 
relation $(x,y) \in f$. Inversion and composition of relations may be simply 
defined \cite{GenTop}.

For particular case of models with discrete time and evolution described by
iteration like \mbox{$\epsilon : x_n \mapsto x_{n+1}$} a ``free state theory'' 
may use only functions like $x_{n+1} = \epsilon(x_n)$, but a relation 
$(x_n,x_{n+1}) \in \epsilon$ already may include a ``free choice.'' 
It is interesting also, that irreversible evolution ({\em cf} Ref.~\cite{tH1})
described by usual function after inversion of time arrow $t \to -t$ may be 
represented only by relation.

A relation may be also described by binary function: $R(x,y) = 1$ for 
$(x,y) \in R$, $R(x,y) = 0$ for $(x,y) \notin R$.
There is some analogue of last definition with probability theory and 
quantum mechanics,  
if instead of the binary function to use real or complex one. 
Conversely, events with unit, 
positive and zero probability define some classes or relations like {\em inevitable} 
(deterministic), {\em possible} (stochastic) and {\em impossible} respectively.%

Quantum mechanics gives rules for calculating of
probabilities like $P(J^2_v{}_\A = {\sf a},\ J^2_w{}_\B = {\sf b})$.
Model with {\em relations} is justified here, because it is suggested in 
\cite{CK} to consider only ``deterministic'' events with $P=1$, 
{\em e.g.}, TWIN axiom may be associated with  
$P(J^2_v{}_\A = {\sf a}| J^2_v{}_\B = {\sf a}) = 1$, 
SPIN axiom  is in agreement with $P(J^2_v{}_\A = 1| J^2_u{}_\A = 0) = 1$
for orthogonal vectors $v \perp u$, {\em etc.} 

Description with relation is not ``free state theory''. Let's denote
$\mathfrak{Q}_{\A\B}(v_\A,{\sf a},w_\B,{\sf b})$ relation describing any possible 
measurements of two spins, {\em i.e.}, 
$P(J^2_v{}_\A = {\sf a}, J^2_v{}_\B = {\sf b}) \neq 0$.
Relation $\mathfrak{Q}_{\A\B}$ may not be expressed in general case as some
function $(v_\A,w_\B) \mapsto ({\sf a},{\sf b})$, because the same values
$(v_\A,w_\B)$ may be related with different $({\sf a},{\sf b})$. 
 
The local version of such a model would be pair of relations
$\mathfrak{Q}_{\A}(v_\A,{\sf a})$ and $\mathfrak{Q}_{\B}(w_\B,{\sf b})$.
It is possible to prove that such relation may not exist for any $v_\A$, $w_\B$
and it is consequence of the Conway-Kochen theorem discussed above.
Really, such relation would define two functions $\JJ_\bot^\A(v)$ and $\JJ_\bot^\B(w)$.
Result of $\JJ_\bot^\A(v)$ is minimal second elements ${\sf a}$ between all pairs 
$(v,{\sf a}) \in \mathfrak{Q}_{\A}$ with the same first element $v$
and $\JJ_\bot^\B(w)$ is an analogue minimum
${\sf b}$ between $(w,{\sf b}) \in \mathfrak{Q}_{\B}$.

$\mathfrak{Q}_{\A}$ and $\mathfrak{Q}_{\B}$ are defined as ``possibility'' relations,
so local functions $\JJ_\bot^{\A}$ and $\JJ_\bot^{\B}$ should also define possible outcomes
of measurement. On the other hand, the proof of Conway-Kochen theorem
recollected above shows, that any local function would not satisfy SPIN and TWIN axiom
for some set of arguments.
But $\JJ_\bot^{\A}$ and $\JJ_\bot^{\B}$ as relations are subset of $\mathfrak{Q}_{\A}$ and 
$\mathfrak{Q}_{\B}$. So some members of $\mathfrak{Q}_{\A}$ and $\mathfrak{Q}_{\B}$ 
are impossible and should have probability zero, but it contradicts to definition
of these relations. 

It is also possible to use maximal values of ${\sf a,b}$ and to introduce functions
$\JJ_\top^{\A}$ and $\JJ_\top^{\B}$. Union of $\JJ_\bot^{\A}$ and $\JJ_\top^{\A}$ is 
$\mathfrak{Q}_{\A}$, because there are only two possible values ${\sf a} = 0,1$.
The same is true for $\mathfrak{Q}_{\B}$ 
$$
\mathfrak{Q}_{\A} = \JJ_\bot^{\A} \cup \JJ_\top^{\A},\quad
\mathfrak{Q}_{\B} = \JJ_\bot^{\B} \cup \JJ_\top^{\B}.
$$
So index of
function can be considered as an additional variable and only nonlocal
consideration may show, which values of such variables are compatible with
laws of quantum mechanics.
 
Roughly speaking, introduction of ``local free will'' does not resolve a 
problem with too rigour axioms and additional
freedom of choice rather intensifies that.

\section{Conway-Kochen Model Without Reduction}
\label{App:CKwor}
\subsection{Quantum Network Model}

If TWIN, SPIN and MIN axiom produce problem with reduction, it is
reasonable to consider idea with resolution of nonlocality problem
in interpretation(s) of quantum mechanics without reduction
similar with developed in \cite{DeuHay,Tip0} and already discussed
briefly in \cite{Sim}.

It is convenient for given frame $F = (x,y,z)$ to consider operator \cite{PerQT}
\begin{equation}
\op{K} \equiv \op{K}^F = \op{J}_x^2 - \op{J}_y^2
\label{opK}
\end{equation}
used to measure values \{$J^2_x{}_\A$, $J^2_y{}_\A$, $J^2_z{}_\A$\}
in single measurement due to equations
\begin{equation}
\op{J}_x^2 = \op\Id - (\op{K}^2-\op{K})/2, \quad
\op{J}_y^2 = \op\Id - (\op{K}^2+\op{K})/2, \quad
\op{J}_z^2 = \op{K}^2.
\label{K2J2xyz}
\end{equation}
Operator $\op{K}^F$ has eigenvalues $\mp 1$ and $0$ with eigenvectors
$\ket{\kappa_1^F}$, $\ket{\kappa_2^F}$ and $\ket{\kappa_3^F}$ 
\[
\op{K}^F\ket{\kappa_1^F} = -\ket{\kappa_1^F},\quad
\op{K}^F\ket{\kappa_2^F} = \ket{\kappa_2^F},\quad
\op{K}^F\ket{\kappa_3^F} = 0.
\] 
In this basis entangled state
used in TWIN axiom may be written as \cite{Sim}\footnote{In \cite{Sim}
is used slightly different operator, but eigenvectors are the same.}
\begin{equation}
 \ket{\Upsilon}=
 \frac{1}{\sqrt{3}}\bigl(\ket{\kappa_1}\ket{\kappa_1}
  +\ket{\kappa_2}\ket{\kappa_2}+\ket{\kappa_3}\ket{\kappa_3}\bigl).
\label{Stwin}
\end{equation}

The \Eq{Stwin} does not depend on frame $F$. If in some frame
we have eigenvectors $\ket{\kappa_i}$, for transition to
other frame described by some rotation matrix $R$
it is possible to write \cite{Sim}
\begin{equation}
F' = R F, \qquad \ket{\kappa'_i} = \sum_{j=1}^3 R_{ij} \ket{\kappa_j}.
\label{RotF}
\end{equation}
Quadratic form $Q(a,b) = a_1 b_1 + a_2 b_2 + a_3 b_3$ is invariant with respect to 
orthogonal transformation $R$ and the same is true for expression \Eq{Stwin}
\begin{equation}
\ket{\kappa_1}\ket{\kappa_1}
  +\ket{\kappa_2}\ket{\kappa_2}+\ket{\kappa_3}\ket{\kappa_3} =
\ket{\kappa'_1}\ket{\kappa'_1}
  +\ket{\kappa'_2}\ket{\kappa'_2}+\ket{\kappa'_3}\ket{\kappa'_3}.
\label{Twinv}
\end{equation}

Invariance of the quadratic form demonstrates also, that two nonlocal measurements 
with the same frame should produce the same value $K^F_\A = K^F_\B \in \{0,\pm 1\}$.
It is reformulation of TWIN axiom with eigenvectors of operator $\op{K}$,
because single value $K$ produces three numbers
$J^2_x{}_\A$, $J^2_y{}_\A$, $J^2_z{}_\A$ (in the given frame)
due to \Eq{K2J2xyz}. 

In initial setup \cite{CK} \B\ has only
one direction $J^2_w{}_\B$ and it corresponds to measurement
of $K^2_\B=J^2_w{}_\B$ in any frame $F = (u,v,w)$ with last  
axis $w$ due to \Eq{K2J2xyz}. In fact, additional detailing
due to possibility of measuring all three axes by Bob 
observer produce some subtleties and distinctions with discussed
symmetric model.

It should be mentioned, that entangled state \Eq{Stwin}
is invariant only with respect to SO$(3)$ subgroup of SU$(3)$ group of all
possible local quantum transformations. So, there is some 
difference with entangled Bell state 
$(\ket{0}\ket{1} - \ket{1}\ket{0})/\sqrt{2}$
invariant with respect to any pair of local transformations SU$(2)$ due to 
invariance of simplectic form $a \wedge b = a_1 b_0 - a_0 b_1$ with respect to
any transformations with unit determinant.

Let us now consider different constructions of measurement gates,
appropriate to description of a TWIN-SPIN scheme \Fig{loccomp} without reduction.
It is possble to denote frame chosen by Alice as 
$\ket{\kappa_1^\A}$, $\ket{\kappa_2^\A}$ and $\ket{\kappa_3^\A}$ 
and corresponding projector operators 
$\Proj_j^\A = \ket{\kappa_j^\A}\bra{\kappa_j^\A}$. It is convenient
also first to use for Bob similar scheme with three one-dimensional
projectors $\Proj_j^\B$ instead of one two-dimensional projector
$\Proj_w^\B$ describing measurement of $\op{J}^2_w{}_\B$,   
where $\Proj_3^\B = \op\Id - \Proj_w^\B$ ({\em cf} \cite{Sim}).

In scheme with reduction the measurements in $\op{K^\A}$ or $\op{K^\B}$
basis should produce three classical outcomes $\mp 1$, $0$, that may be 
decoded for simplicity in three numbers $j = 1,2,3$ and transferred
to some place for comparison. 

Measurement without reduction should transfer data on some auxiliary 
quantum state $\ket{j}$ \Fig{loccomp} instead of production of classical outcome
and due to impossibility to clone quantum state there are two
basic scheme. First one is generalisation of {\em qubit measurement
gate} already used earlier in quantum networks model for resolution 
of quantum nonlocality problem \cite{DeuHay}. 

For qutrit (spin-1) case and notation used here {\em measurement gate} 
works with measured system and auxiliary carrier in initial state $\ket{0}$. 
It is also more convenient to use here
zero-based indexes $j = 0,1,2$ for bases $\ket{\kappa_j^F}$ and 
projectors $\Proj_j^F$ for some frame $F=(x,y,z)$. For this frame
measurement gate acts on basic states as
\begin{equation}
 \Meas^F \bigl(\ket{\kappa_j^F}\ket{k}\bigr) = \ket{\kappa_j^F}\ket{k+j \mod 3},
 \quad
 \Meas^F : \ket{\kappa_j^F}\ket{0} \mapsto \ket{\kappa_j^F}\ket{j}
\label{MFact}
\end{equation}
and may be represented as
\begin{equation}
 \Meas^F = \sum_{j=0}^2 {\Proj_j^F \otimes \op{U}^j},
\label{MFrep}
\end{equation}
where $\op{U}^j$ is power of cyclic shift,\footnote{Formally, instead of 
$\op{U}^j$ also may be used any unitary operator with property
$\ket{0} \mapsto \ket{j}$.} defined on basis vectors as 
$\ket{k} \mapsto \ket{k+j \mod 3}$. 

Yet another gate convenient for nondestructive measurement is
SWAP (exchange) gate $\Swap$
\begin{equation}
 \Swap^F \bigl(\ket{\kappa_j^F}\ket{k}\bigr) = \ket{\kappa_k^F}\ket{j}, 
 \quad
 \Swap^F : \ket{\kappa_j^F}\ket{0} \mapsto \ket{\kappa_0^F}\ket{j}, 
\label{SFact}
\end{equation}
represented as
\begin{equation}
 \Swap^F = 
 \sum_{j,k=0}^2 {\ket{\kappa_k^F}\bra{\kappa_j^F} \otimes \ket{j}\bra{k}}.
\label{SFrep}
\end{equation}
Advantage of SWAP gate is because of system in any initial state is not entangled
with carrier after operation
\begin{equation}
 \Swap^F : \Bigl(\sum_{j=0}^2\alpha_j\ket{\kappa_j^F}\Bigr)\ket{0} 
 \mapsto \ket{\kappa_0^F}\sum_{j=0}^2\alpha_j\ket{j},
\label{MFnent}
\end{equation}
unlike the measurement gate
\begin{equation}
 \Meas^F : \Bigl(\sum_{j=0}^2\alpha_j\ket{\kappa_j^F}\Bigr)\ket{0} 
 \mapsto \sum_{j=0}^2\alpha_j\ket{\kappa_j^F}\ket{j}.
\label{MFent}
\end{equation}

For work with entangled system it is necessary to have two carriers and
apply two gates $\Meas^\A \otimes \Meas^\B$ or $\Swap^\A \otimes \Swap^\B$.
For consideration of situation described in TWIN axiom it is enough
to consider choice of same frame $F$ by Alice and Bob. State $\ket{\Upsilon}$
\Eq{Stwin} may be written in the same form for any frame due to
invariance property \Eq{Twinv}. With two carriers it may be written as
\begin{equation}
 \Ket{\Upsilon}_{1,3}\Ket{0}_2\Ket{0}_4=
 \frac{1}{\sqrt{3}}\Bigl(\Ket{\kappa_0^F}_1\Ket{\kappa_0^F}_3
  +\Ket{\kappa_1^F}_1\Ket{\kappa_1^F}_3+\Ket{\kappa_2^F}_1\Ket{\kappa_2^F}_3\Bigr)
  \Ket{0}_2\Ket{0}_4,
\label{Stwin00}
\end{equation}
where lower index is number of system, {\em i.e.}, 1 --- Alice's subsystem, 
2 --- Alice's carrier, 3 --- Bob's subsystem, 4 --- Bob's carrier,
and 1,3 --- whole entangled system.

After application of two measurement gates to \Eq{Stwin00} we have
entangled state of system and carriers
\begin{equation}
  \Meas_\A^F \otimes \Meas_\B^F : \Ket{\Upsilon}_{1,3}\Ket{0}_2\Ket{0}_4 \mapsto
  \frac{1}{\sqrt{3}}\Bigl(\Ket{\kappa_0^F}_1\Ket{0}_2\Ket{\kappa_0^F}_3\Ket{0}_4
  +\Ket{\kappa_1^F}_1\Ket{1}_2\Ket{\kappa_1^F}_3\Ket{1}_4+
  \Ket{\kappa_2^F}_1\Ket{2}_2\Ket{\kappa_2^F}_3\Ket{2}_4\Bigr).
\label{MMtwin}
\end{equation}
After receiving of carriers it is necessary 
to apply some operation to states of carriers, say {\em comparison gate} 
\begin{equation}
\op{\mathtt C} = \Meas^* : \ket{k}\ket{j} \mapsto \ket{k}\ket{k-j \mod 3},
\label{GCmp}
\end{equation}
denoted as {\tt C} on \Fig{loccomp}. After such operation state \Eq{MMtwin}
is transformed to
\begin{equation}
  \Ket{\Xi^F_+} \equiv
  \frac{1}{\sqrt{3}}\Bigl(\Ket{\kappa_0^F}_1\Ket{\kappa_0^F}_3\Ket{0}_2
  +\Ket{\kappa_1^F}_1\Ket{\kappa_1^F}_3\Ket{1}_2+
  \Ket{\kappa_2^F}_1\Ket{\kappa_2^F}_3\Ket{2}_2\Bigr)\Ket{0}_4,
\label{MMCtwin}
\end{equation}
{\em i.e.}, represents product of some entangled state and fourth
``comparison state.'' Zero value of this state ensures successful
comparison. 

Application of two SWAP gates produces simpler expression instead of \Eq{MMtwin}
\begin{equation}
  \Swap_\A^F \otimes \Swap_\B^F : \Ket{\Upsilon}_{1,3}\Ket{0}_2\Ket{0}_4 \mapsto
  \Ket{\kappa_0^F}_1\Ket{\kappa_0^F}_3\frac{1}{\sqrt{3}}\bigl(\Ket{0}_2\Ket{0}_4
  +\Ket{1}_2\Ket{1}_4+\Ket{2}_2\Ket{2}_4\bigr).
\label{XXtwin}
\end{equation}
Composition of comparison gate $\Meas^*$ \Eq{GCmp} to carriers-$\scriptstyle 2,4$ and
inverse Fourier transform to carrier-$\scriptstyle 2$ reduces \Eq{XXtwin} to simplest
analogue of \Eq{MMCtwin}
\begin{equation}
  \Ket{\Xi^{F}_0} \equiv
  \Ket{\kappa_0^F}_1\Ket{\kappa_0^F}_3\Ket{0}_2\Ket{0}_4.
\label{XXCFtwin}
\end{equation}


\paragraph{Note:}\label{NoteAX}
It may be argued, that such symmetric experiments with measurement of the same 
values for the same frame formally could be reproduced with a hidden variable 
model, if to assign to each frame a triple of values $\{0,1,2\}$.
But such a scheme works only for axes in the frames are in the same order.
A scheme suggested in \cite{CK} may not rely on such order and any 
nonlocal models are impossible

Lets's consider the initial (asymmetric) setup suggested in \cite{CK}, 
when \B\ measures $J^2_w{}_\B$ only for one direction $w$. 
In such a case carrier may have only two states and measurement 
gate may be described as
\begin{equation}
 \Meas^w \equiv \Proj^w \otimes \bigl(\ket{0}\bra{0}+\ket{1}\bra{1}\bigr)
  + (\op\Id - \Proj^w) \otimes \bigl(\ket{0}\bra{1}+\ket{1}\bra{0}\bigr).
\label{Mwrep}
\end{equation}
Let us consider general case, when $w$ may be decomposed using axes $x, y, z$ in given
frame $F$ as 
\begin{equation}
 w = w_0 x + w_1 y + w_2 z, \quad
 \ket{\kappa^w} = w_0 \ket{\kappa_0^F} + w_1 \ket{\kappa_1^F} + w_2 \ket{\kappa_2^F}. 
\label{wjkj}
\end{equation}
Now application of modified ``$3 \times 2$'' measurement gates produces
\begin{equation}
  \Meas_\A^F \otimes \Meas_\B^w : \Ket{\Upsilon}_{1,3}\Ket{0}_2\Ket{0}_4 \mapsto
  \frac{1}{\sqrt{3}}\sum_{j=0}^2\ket{\kappa_j^F}\ket{j}
  \Bigl(w_j\ket{\kappa^w}\ket{0} + 
    \bigl(\ket{\kappa_j^F} - w_j\ket{\kappa^w}\bigr)\ket{1}\Bigr)
\label{MMwtwin}
\end{equation}
and for particular cases $w=x$, $w=y$ and $w=z$ \Eq{MMwtwin} produces
\begin{subequations}
\begin{align}
  \Meas_\A^F \otimes \Meas_\B^x : \Ket{\Upsilon}_{1,3}\Ket{0}_2\Ket{0}_4 &\mapsto
  \frac{1}{\sqrt{3}}\Bigl(\Ket{\kappa_0^F}_1\Ket{0}_2\Ket{\kappa_0^F}_3\Ket{0}_4
  +\Ket{\kappa_1^F}_1\Ket{1}_2\Ket{\kappa_1^F}_3\Ket{1}_4+
  \Ket{\kappa_2^F}_1\Ket{2}_2\Ket{\kappa_2^F}_3\Ket{1}_4\Bigr),\\
  \Meas_\A^F \otimes \Meas_\B^y : \Ket{\Upsilon}_{1,3}\Ket{0}_2\Ket{0}_4 &\mapsto
  \frac{1}{\sqrt{3}}\Bigl(\Ket{\kappa_0^F}_1\Ket{0}_2\Ket{\kappa_0^F}_3\Ket{1}_4
  +\Ket{\kappa_1^F}_1\Ket{1}_2\Ket{\kappa_1^F}_3\Ket{0}_4+
  \Ket{\kappa_2^F}_1\Ket{2}_2\Ket{\kappa_2^F}_3\Ket{1}_4\Bigr),\\
  \Meas_\A^F \otimes \Meas_\B^z : \Ket{\Upsilon}_{1,3}\Ket{0}_2\Ket{0}_4 &\mapsto
  \frac{1}{\sqrt{3}}\Bigl(\Ket{\kappa_0^F}_1\Ket{0}_2\Ket{\kappa_0^F}_3\Ket{1}_4
  +\Ket{\kappa_1^F}_1\Ket{1}_2\Ket{\kappa_1^F}_3\Ket{1}_4+
  \Ket{\kappa_2^F}_1\Ket{2}_2\Ket{\kappa_2^F}_3\Ket{0}_4\Bigr).
\end{align}
\label{MMwtwin3}
\end{subequations}
These expressions are analogues of \Eq{MMtwin}. So despite of importance of
such modification for impossibility of hidden variable model such asymmetric 
experiment, it is very close to symmetric one. 

An essential property of vector $w$ is possibility to be equal with {\em any} between 
three axis. 

Let us denote for frame $F = (x,y,z)$ and vector $w$ 
\begin{equation}
 n^{F,w}=
 {\renewcommand{\arraystretch}{0.75}
 \left\{
 \begin{array}{ll}
  0, & w = x\\  1, & w = y\\ 2, & w = z\\
  -1, & \text{otherwise}
 \end{array}
 \right.
 },\qquad
 a \comp b  \equiv  \begin{cases}
                  0, & a = b \\
                  1, & a \ne b 
                \end{cases},  
\label{nFw}
\end{equation}
then \Eq{MMwtwin3} for $w=x$, $w=y$ or $w=z$ may be rewritten
\begin{equation}
  \Meas_\A^F \otimes \Meas_\B^w : \Ket{\Upsilon}_{1,3}\Ket{0}_2\Ket{0}_4 \mapsto
  \frac{1}{\sqrt{3}}\sum_{j=0}^2\ket{\kappa_j^F}\ket{j}\ket{\kappa_j^F}\ket{ n^{F,w}\comp j},
\quad (w = x) \lor (w = y) \lor (x = z) 
\label{MMwtwin1}
\end{equation}
and comparison operator for \Eq{MMwtwin3},  \Eq{MMwtwin1} should have
more difficult form than \Eq{GCmp}.
New version of operation of comparison should depend on frame
and may be written as
\begin{equation}
 \ket{c_1}\ket{c_2}\ket{c} \mapsto
  \ket{c_1}\ket{c_2}\Ket{n^{F,w} \comp c_1 \comp c_2 \comp c}, 
\label{GFcmp}
\end{equation}
where  $\ket{c_1}$,  $\ket{c_2}$  are states of carriers and $\ket{c}$ is
auxiliary {\em comparison qubit} with initial state is always $\ket{0}$
and final state $\ket{0}$ only for successful comparison.%
\footnote{Operation \Eq{GFcmp} is unitary because for qubit $a \comp b = a + b \pmod 2$
and, so, it is ``quantum function evaluation'' $\ket{c + f(c_1,c_2) \mod 2}$
for $f(c_1,c_2) = n^{F,w} \comp c_1 \comp c_2$.}

\subsection{Classical Equipment}

Considered model produces questions about relation with {\em real} experiments
on quantum nonlocality. It may be objected, that instead of quantum carriers
are still used classical wires and coincidence schemes \cite{Asp} and it usually 
does not resemble quantum comparison gates discussed above.

Minimal realistic modification ---
is description of signals with many elementary carriers (modes), {\em e.g.},
$\ket{\mbs 0} \equiv \ket{0}\ket{0} \ldots \ket{0}$ instead of $\ket{0}$, 
{\em etc.} Formally, for $N \to \infty$ it may be compared with carriers, 
described by continuous quantum variables. For continuous quantum variables
basic states may be naturally associated with classical parameters. Here the operation of 
comparison still may be written directly 
\begin{equation}
\op{\mathbf C} : 
\ket{\mbs k}\ket{\mbs j} \mapsto \ket{\mbs k}\ket{\mbs{k-j}}
\label{bfGCmp}
\end{equation}
and in classical limit it corresponds to understanding formula
like $(x,y) \mapsto x-y$, there $x$ and $y$ are some classical
values, {\em e.g.}, pointers positions of some devices.  

If to save notation $\ket{\mbs k}$ for a state of some set of carriers, even
for simpler case with SWAP gates an analogue of \Eq{XXtwin} generates couple of 
signals in a ``cat state'' 
\begin{equation}
  \Swap_\A^F \otimes \Swap_\B^F : 
  \Ket{\Upsilon}_{1,3}\Ket{\mbs 0}_2\Ket{\mbs 0}_4 \mapsto
  \Ket{\kappa_0^F}_1\Ket{\kappa_0^F}_3
  \frac{1}{\sqrt{3}}\bigl(\Ket{\mbs 0}_2\Ket{\mbs 0}_4
  +\Ket{\mbs 1}_2\Ket{\mbs 1}_4+\Ket{\mbs 2}_2\Ket{\mbs 2}_4\bigr).
\tag{\ref{XXtwin}$'$}
\label{bfXXtwin}
\end{equation}
Analogue of \Eq{MMtwin} is even more cumbersome
\begin{equation}
  \Meas_\A^F \otimes \Meas_\B^F : 
  \Ket{\Upsilon}_{1,3}\Ket{\mbs 0}_2\Ket{\mbs 0}_4 \mapsto
  \frac{1}{\sqrt{3}}\Bigl(
   \Ket{\kappa_0^F}_1\Ket{\kappa_0^F}_3\Ket{\mbs 0}_2\Ket{\mbs 0}_4
  +\Ket{\kappa_1^F}_1\Ket{\kappa_1^F}_3\Ket{\mbs 1}_2\Ket{\mbs 1}_4
  +\Ket{\kappa_2^F}_1\Ket{\kappa_2^F}_3\Ket{\mbs 2}_2\Ket{\mbs 2}_4\Bigr).
\tag{\ref{MMtwin}$'$}
\label{bfMMtwin}
\end{equation}

In ``consequent'' Everettian interpretation \Eq{bfMMtwin} and \Eq{bfXXtwin} 
formally do not need for some {\em explicit} operation of comparison, because  
states of carriers are almost ``automatically'' associated with one of three 
posible combinations
\begin{equation}
\Ket{\kappa_k^F}_1\Ket{\kappa_k^F}_3\Ket{\mbs k}_2\Ket{\mbs k}_4,
\quad k=0,1,2,
\label{Mtwin}
\end{equation}
due to usual arguments of relative states formulation \cite{Ev}. The same is true
for more complicated analogue of \Eq{MMwtwin3} or \Eq{MMwtwin1}
\begin{equation}
\Ket{\kappa_k^F}_1\Ket{\kappa^w}_3\Ket{\mbs k}_2\Ket{\mbs n}_4,
\quad k=0,1,2, \quad n = 0,1,
\label{Mwtwin}
\end{equation}
with $n = n^{F,w} \comp k$, if $w =x$, $w=y$ or $w=z$, {\em cf\/} \Eq{nFw}.

Considered approach may be discussed also if reduction is only delayed till a
final comparison after arriving of two signals in the same place. For such a case
problem of nonlocality is formally avoided, but instead of problem of nonlocal 
collapse for two simple quantum systems here appears questionable effect of reduction 
of ``cat states'' with huge amount of elementary carriers.

\section{Example of Decomposition}
\label{App:Decomp}

Here is represented a decomposition of density matrix 
$\ket{\Upsilon}\bra{\Upsilon}$ for 
$\Upsilon = (\ket{1}\ket{1}+\ket{2}\ket{2}+\ket{3}\ket{3})/\sqrt{3}$
as some combination of product states with positive and negative
coefficients.

Let us introduce nine states
\begin{align}
\ket{\xi_1} &\equiv \ket{1},& 
\ket{\xi_2} &\equiv \ket{2},& 
\ket{\xi_3} &\equiv \ket{3}, \nonumber\\
\ket{\xi_4} &= (\ket{2}+\ket{3})/\sqrt{2}, &
\ket{\xi_5} &= (\ket{3}+\ket{1})/\sqrt{2}, &
\ket{\xi_6} &= (\ket{1}+\ket{2})/\sqrt{2}, \\
\ket{\xi_7} &= (\ket{2}+i\ket{3})/\sqrt{2}, &
\ket{\xi_8} &= (\ket{3}+i\ket{1})/\sqrt{2}, &
\ket{\xi_9} &= (\ket{1}+i\ket{2})/\sqrt{2}. \nonumber
\end{align}
Density matrixes for these states
\begin{equation}
 \op{\rho}_k \equiv \ket{\xi_k}\bra{\xi_k}, 
 \qquad k = 1,\dots,9.
\end{equation}
produce a basis in space of Hermitian $3 \times 3$ matrixes and 81 tensor products 
$\rho_k \otimes \rho_j$ is a basis in the tensor product of two such spaces \cite{VlPr}.
Coefficients $c_{kj}$ of decomposition for composite system 
$\ket{\Upsilon}\bra{\Upsilon} = \sum_{k,j=1}^9 c_{kj} \rho_k \otimes \rho_j$ are
represented in table below and may be found using a standard methods for solution 
of a linear system (with 81 equations).
\begin{equation}
\begin{array}{c|rrr|rrr|rrr}
\otimes & \op{\rho}_1 & \op{\rho}_2 & \op{\rho}_3 & 
\op{\rho}_4 & \op{\rho}_5 & \op{\rho}_6 &
\op{\rho}_7 & \op{\rho}_8 & \op{\rho}_9 \\ \hline
\op{\rho}_1 &
{1 / 3} &  &  &  & -{1 / 3} & 
-{1 / 3} &  & {1 / 3} & {1 / 3}\\
\op{\rho}_2 & 
 & {1 / 3} &  & -{1 / 3} &  & 
-{1 / 3} & {1 / 3} &  & {1 / 3} \\
\op{\rho}_3 &
 &  & {1 / 3} & -{1 / 3} & -{1 / 3} & 
 & {1 / 3} & {1 / 3} &  \\ \hline
\op{\rho}_4 & 
  & -{1 / 3} & -{1 / 3} & {2 / 3} &  &
 &  &  &  \\ 
\op{\rho}_5 & 
-{1 / 3} &   & -{1 / 3} &  & {2 / 3} & 
 &  &  &  \\
\op{\rho}_6 &  
-{1 / 3} & -{1 / 3} &  &  &  &
{2 / 3} &  &  & \\ \hline
\op{\rho}_7 &
 &{1 / 3} & {1 / 3} &  &  &
 & -{2 / 3} &  &  \\
\op{\rho}_8 &
{1 / 3} &   & {1 / 3} &  &  & 
 &  & -{2 / 3} &  \\
\op{\rho}_9 & 
{1 / 3} & {1 / 3} &  &  &  &
 &  &  & -{2 / 3}  
\end{array}
\label{Decomp81}
\end{equation}

\newpart


\end{document}